\documentclass{emulateapj}
\usepackage{psfig}
\usepackage{bm}
\usepackage{amsmath}

\newcommand{\eqn}[1]{equation~(\ref{#1})}
\newcommand{\fig}[1]{Figure~\ref{#1}}
\newcommand{\tab}[1]{Table~\ref{#1}}

\newcommand{\bvec}[1]{\boldsymbol{\mathrm{#1}}}
\newcommand{\dif}{\ensuremath{\mathrm{d}}}

\newcommand{\jwst}{{\it JWST}}                   
\newcommand{\hst}{{\it HST}}                     

\newcommand{\htwoo}{H$_2$O}
\newcommand{\co}{CO}
\newcommand{\chfour}{CH$_4$}

\newcommand{\nfields}{11}

\shorttitle{Disk Brown Dwarfs in \jwst\ Surveys}
\shortauthors{Ryan \& Reid}
\begin{document}

\title{The Surface Densities of Disk Brown Dwarfs in {\it JWST} Surveys}
\email{rryan@stsci.edu}

\author{R. E. Ryan Jr.\altaffilmark{1},
I.~N.~Reid\altaffilmark{1}}
\altaffiltext{1}{Space Telescope Science Institute, 3700 San Martin Dr., Baltimore, MD 21218}

\begin{abstract}

We  present predictions for  the surface  density of  ultracool dwarfs
(with spectral types M8--T8) for a host of deep fields that are likely
to be  observed with the {\it  James Webb Space  Telescope}.  Based on
simple  thin and  thick/thin disk  (exponential) models,  we  show the
typical  distance modulus  is $\mu\!\approx\!9.8$~mag,  which  at high
Galactic  latitude  is   $5\log{(2\,z_{scl})}-5$.   Since  this  is  a
property of  the density  distribution of an  exponential disk,  it is
independent of  spectral type or stellar sample.   Using the published
estimates  of the ultracool  dwarf luminosity  function, we  show that
their  number counts  typically peak  around $J\!\sim\!24$~mag  with a
total surface density of $\Sigma\!\sim\!0.3$~arcmin$^{-2}$, but with a
strong dependence on galactic  coordinate and spectral type.  Owing to
the exponential shape of the  disk, the ultracool dwarfs are very rare
at  faint magnitudes  ($J\!\geq\!27$~mag), with  typical  densities of
$\Sigma\!\sim\!0.005$~arcmin$^{-2}$  (or  $\sim\!20$\%  of  the  total
contribution within the field).  Therefore in the very narrow and deep
fields, we  predict there are only  a few ultracool  dwarfs, and hence
these  stars are  likely  not  a severe  contaminant  in searches  for
high-redshift galaxies.  Furthermore the ultracool dwarfs are expected
to  be  considerably  brighter  than the  high-redshift  galaxies,  so
samples near the faint-end of the high-redshift galaxy population will
be the  purest.  We present  the star-count formalism in  a simplified
way so that observers may easily predict the number of stars for their
conditions (field, depth, wavelength, etc.).

\end{abstract}


\section{Introduction} \label{sec:intro}

Number-magnitude counts  have a long tradition of  providing a simple,
but effective  tool for probing  the distribution of stars  within the
Milky Way and the large-scale structure of the Universe \citep[such as
  visual star-gaging of][]{herschel}.   The application of photography
to  astronomical  research in  the  late  1800s  provided a  means  of
obtaining   permanent  records  of   celestial  objects,   leading  to
systematic  surveys such  as the  Plan of  Areas \citep{kapt}  and the
star-count  analyses  \citep{seares,bok}.   Additionally,  photography
also opened the  way for surveys of the  distribution of extragalactic
nebulae,   notably   the  Lick   survey   by   \citet{sw54}  and   the
\citet{abell58}   catalog  of  galaxy   clusters,  derived   from  the
wide-field Schmidt plates of the first Palomar Observatory Sky Survey.

Photography  captured images  of  the sky  for  posterity, but  visual
inspection of  plates can only yield a  qualitative understanding. The
development of scanning densitometers  in the 1970s provided the first
means of quantifying photographic data for statistical analyses.  When
combined  with  deep  imaging  with  4-meter  telescopes,  photography
yielded the  first reliable color-magnitude diagrams  for field stars
and galaxies fainter than $\sim\!20$~mag \citep{kron80}, and automated
scans of  wide-field Schmidt plate mapped the  stellar distribution at
bright    magnitudes    \citep[e.g.][]{hewett81,gilm85},    but    see
\citet{reid93,chin13}  for  a   more  extensive  discussion  of  those
developments. In  more recent  years, these investigations  have given
way  to direct  digital  imaging, whether  through narrow  pencil-beam
surveys,    as    exemplified     by    the    Hubble    Deep    Field
\citep[HDF;][]{williams96} and its successors, or through near all-sky
surveys, such as the Sloan Digital Sky Survey \citep[SDSS;][]{york00}.

By and  large, stars and  galaxies occupy distinct domains  in imaging
surveys.   At  high galactic  latitudes,  stars  are typically  bright
($V\!\sim\!20$~mag),  while galaxies  dominate  at fainter  magnitudes
($v\!\gtrsim\!25$~mag).    However   rare   objects  often   challenge
conventional models  and highlight short-comings  in current theories.
For  example,   \citet{gilm81}  show   that  rare,  faint   stars  can
significantly contaminate samples of distant objects. In recent years,
the focus has moved to significantly higher redshifts with the deepest
surveys,  aided  partly  by  gravitational  lensing,  reaching  beyond
$z\!\sim\!10$  \citep{coe13,ellis13}.  At  those redshifts,  the Lyman
break  moves longwards  of $\lambda\!\sim\!1~\mu$m  and  galaxies have
extremely    red   colors    at    near-infrared   wavelengths.     As
\citet{wilkins14}  have pointed out,  very high-redshift  galaxies and
ultracool  dwarfs often  have similar  near-infrared colors.   In this
paper,  we   consider  the   likely  surface  density   of  MLT-dwarfs
\citep{kirk99,cush11}   and  their  potential   to  bias   studies  of
high-redshift galaxies.

This paper is organized as follows:~In \S~\ref{sec:background} we give
a  brief discussion  of  stellar  populations in  the  Milky Way,  the
relevant  properties  of ultracool  dwarfs,  and  our  choice of  {\it
  representative} deep fields.   In \S~\ref{sec:surveys} we detail the
star-count  formalism.   In   \S~\ref{sec:discuss}  we  discuss  these
results  in the  context of  deep fields  with the  {\it  Hubble Space
  Telescope} and  the {\it James  Webb Space Telescope}.   Finally, in
\S~\ref{sec:summary} we  conclude with a brief  summary, reviewing the
key points.  Throughout  this paper, we take care  to explicitly state
the  magnitude system,  to avoid  confusion between  Vega-based (often
used in the Galactic and  stellar community) and AB-based (the {\it de
  facto} standard in extragalactic work) magnitudes.

\section{Background}\label{sec:background}

\subsection{The Stellar Populations of the Milky Way}\label{sec:pops}

The  resolved stellar  constituents of  nearby galaxies  are generally
categorized   as   members   of   distinct   populations.    Following
\citet{baade44}  and  \citet{oort58}, a  {\it  stellar population}  is
characterized  as a collection  of stars  that have  similar dynamical
properties and share a  common evolutionary scenario. Within the Milky
Way,  the main  populations are  the thin  disk, the  thick  disk, the
stellar halo, and the Galactic bulge/bar. The last-named population is
generally confined within the  central regions of the Galaxy (although
radial  migration  may lead  to  some  local  representation), and  we
therefore focus  the first two populations as  representative of stars
in  the  outer  regions  of  the  Galaxy  in  general  and  the  Solar
Neighborhood in  particular. The main properties  of these populations
are well summarized by \citet{freeman12}.

Considering the three local populations, the thin disk is the dominant
baryonic component,  with a total  mass of $\sim\!5\times10^9~M_\odot$
and  encompassing  the gas  and  dust  contributing  the current  star
formation. The density distribution is generally well represented by a
double-exponential,     with      a     radial     scale-length     of
$\sim\!2.5-2.7$~kpc. Gas,  dust, and young stars  are closely confined
to the  Galactic midplane,  with the vertical  distribution increasing
rapidly  with age  as  the  velocity increases  due  to scattering  by
massive objects such  as molecular clouds \citep{spitzer,wielen}.  The
oldest stars in the disk have ages $\sim\!8-10$~Gyr and distributed in
a disk of scale-height of $\sim\!250$~pc \citep{juric08}.

The  thick  disk is  a  more  extended  component, again  following  a
double-exponential  distribution with  a radial  scale-length  that is
similar to the thin disk. Originally identified within the Galaxy from
star count analyses by  \citet{gr83}, the vertical distribution can be
matched    with   a    scale-height   of    $\sim\!800-900$~pc.    The
color-magnitude  diagram  clearly  indicates   that  this  is  an  old
population  ($\sim\!10-12$~Gyr),  with  essentially no  on-going  star
formation.    While  the  exact   origin  remains   unclear,  detailed
spectroscopic  analyses  show  that  thick-disk  stars  have  enhanced
abundances of $\alpha$-elements  \citep{bensby13}, indicating that the
population   formed   rapidly,  before   type   Ia  supernovae   could
significantly  enhance  the  iron  abundance.  The  local  density  of
thick-disk  stars is  $\sim\!8-10$\% that  of  the thin  disk, with  a
likely total mass of $\sim\!(1-2)\times 10^9~M_\odot$.

\subsection{Colors of the Ultracool Dwarf Population}\label{sec:ucds}

\begin{figure}
\epsscale{1.2}
\plotone{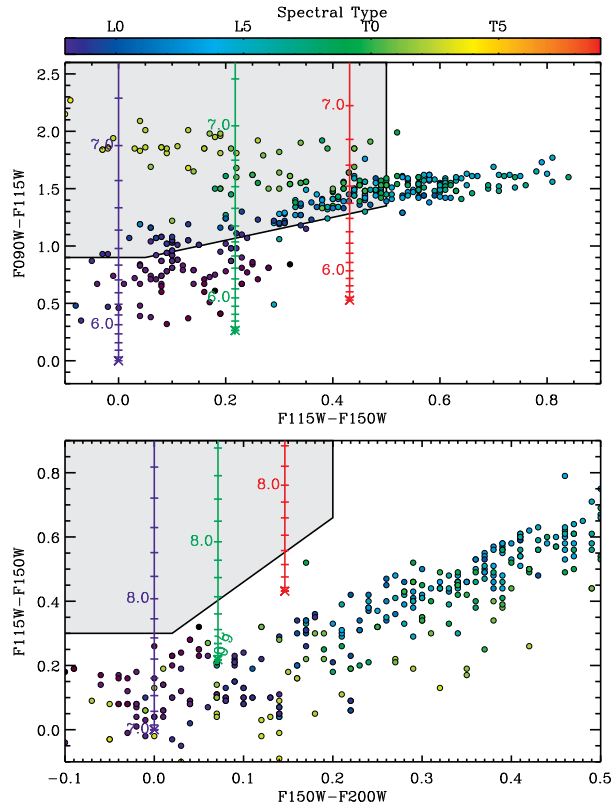}
\caption{\jwst/NIRCam  broadband colors.   Each  point represents  one
  object from  the IRTF/SpeX library  of A.~Burgasser where  the color
  encodes spectral  type as indicated  by the color bar.   The colored
  tracks show  the path  of idealized, power-law  spectra of  the form
  $f_\lambda\!\propto\!\lambda^{-\beta}$  with  $\beta\!=!-2$  (blue),
  $\beta\!=\!-1$  (green), and  $\beta\!=\!0$ (red),  where  each tick
  represents  $\Delta z\!=\!0.1$.   The  gray region  shows a  typical
  color-color    region   for    selecting    high-redshift   galaxies
  ($z\!\gtrsim\!6.5$),   which   demonstrates   the  well-known   {\it
    identification  degeneracy} widely  discussed  in the  literature.
  While  this   degeneracy  is  readily  broken  with   the  suite  of
  medium-bands designed  to sample molecular  absorption (from \htwoo,
  \co,     and     \chfour)     or    additional     broadbands     at
  $\lambda\!\gtrsim\!2~\mu$m,  such data  are  rarely available.   The
  stars  (unlike  the high-redshift  galaxies)  are not  homogeneously
  distributed on  the sky and so  we provide simple  models to predict
  the      surface     density      of      Galactic     stars      in
  \S~\ref{sec:surveys}. \label{fig:colors}}
\end{figure}

Ultracool dwarfs have effective temperatures $T\!\lesssim\!2500$~K and
emergent spectra characterized by  absorption from broad molecular and
narrow  resonance   features  (e.g.~H$_2$O,  FeH,   TiO,  CO,  CH$_4$,
\ion{Na}{1}, and \ion{K}{1}).   Consequently these low-mass stars have
very red optical/near-infrared broad-band colors, which are similar to
galaxies at  $z\!\gtrsim\!6$ \citep{yan03,ryan05,cab08,wilkins14}.  In
\fig{fig:colors},   we   show   the  \jwst/NIRCam   broadband   colors
synthesized  from   the  IRTF/SpeX  library\footnote{as   compiled  by
  A.~Burgasser              and              available              at
  http://pono.ucsd.edu/~adam/browndwarfs/spexprism/}     as    colored
points.  The  colored lines show  the tracks of powerlaw  spectra with
$f_\lambda\!\propto\!\lambda^{-\beta}$   with   blue  ($\beta\!=\!2$),
green ($\beta\!=\!1$), and red ($\beta\!=\!0$).  Additionally, we show
an  example selection region  for $z\!\gtrsim\!7$  galaxies as  a gray
polygon.  With  a limited number of broad-bands,  the ultracool dwarfs
and  high-redshift  galaxies  have   similar  colors  and  are  easily
misidentified.  Medium-bands tailored to sample the molecular features
in  the  low-mass stars  can  easily  break  this {\it  identification
  degeneracy},  but in  their absence  it is  important to  assess the
potential number of stars in the sample.

\subsection{Stellar Luminosities}\label{sec:lums}

\begin{figure}
\epsscale{1.2} 
\plotone{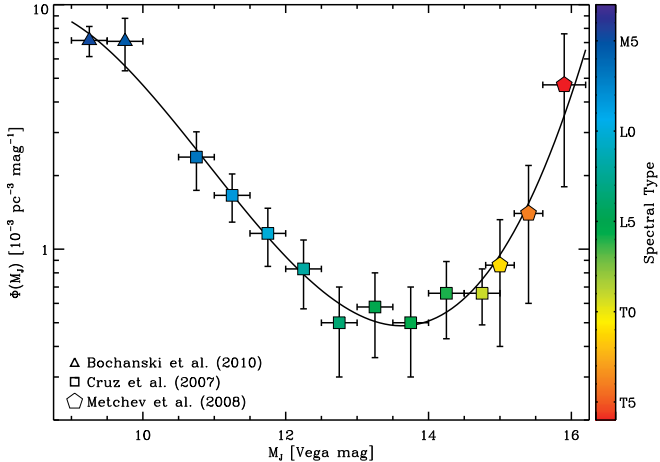}
\caption{$J({\rm  Vega})$-band luminosity  function.  The  symbols and
  colors represent  the author from which  the data are  taken and the
  spectral type, respectively.  The  solid line shows the fourth-order
  fit  given  in  \eqn{eqn:polylf}.  To  map  between  luminosity  and
  spectral type,  we adopt the \citet{hawl02}  relation.  As mentioned
  in the text, we do not attempt to propagate the uncertainty on these
  data (or  the uncertainty on the polynomial  fit) into uncertainties
  on  the observed  number  counts,  as our  primary  goal is  provide
  estimates for total number of brown dwarfs.  \label{fig:lf}}
\end{figure}

We adopt  the local $J$-band luminosity function  pieced together from
\citet{cruz07,boch10,metchev},   but  a   general   calculation  could
transform to  an arbitrary near-infrared passband  using the IRTF/SpeX
library.   However these density  estimates have  modest uncertainties
($\delta\Phi/\Phi\!\sim\!30\%$), so we take our luminosity function to
be a fourth-order polynomial fit for $8.5\!\leq\!J\!\leq\!16.5$~mag:
\begin{equation}\label{eqn:polylf}
\begin{aligned}
\log\Phi(J)={} & -0.30+0.11\,(J-14)+0.15\,(J-14)^2\\
               & +0.015\,(J-14)^3-0.00020\,(J-14)^4
\end{aligned}           
\end{equation}
where $\Phi(J)$ has units of $10^{-3}$~pc$^{-3}$~mag$^{-1}$ and $J$ is
Vega-based.   In  \fig{fig:lf},  we  show  our  polynomial  luminosity
function (solid line) and the observations as color symbols, where the
color and  shape represent spectral type  and reference, respectively.
As our primary  goal is to predict the  number counts at high-Galactic
latitude, we  do not  propagate the uncertainty  in these data  or the
corresponding  polynomial  model.  Finally,  we  adopt  the  $M_J({\rm
  Vega})$--spectral type relation of \citet{hawl02}.

\subsection{Representative Deep Fields}\label{sec:fields}
Although the precise details of the forthcoming \jwst\ deep fields are
not yet finalized, a reasonable  strawman plan would use the broadband
filters in the standard fields.  We consider \nfields~fields, the five
fields  from  the  Cosmic  Assembly Near-infrared  Deep  Extragalactic
Legacy Survey \citep[CANDELS;][]{amk,nag}  and the six Hubble Frontier
Fields\footnote{http://www.stsci.edu/hst/campaigns/frontier-fields/}
\citep[HFF; Lotz  et al.~in prep][]{lotz14}.   In \tab{tab:fields}, we
list the observational properties of these \nfields~fields, but regard
COSMOS,   Abell   S1063,   Abell   2744,  and   MACS~J0717   as   {\it
  representative}  fields,  since  they  cover a  range  of  celestial
coordinates and give the largest spread in the following predictions.

\begin{table*}
\caption{Representative \jwst\ Fields}
\label{tab:fields}
\begin{tabular*}{\textwidth}
    {@{\extracolsep{\fill}}lccccc}
\hline\hline
\multicolumn{1}{c}{Name} & \multicolumn{1}{c}{$\alpha$} & \multicolumn{1}{c}{$\delta$} & \multicolumn{1}{c}{$\ell$} & \multicolumn{1}{c}{$b$} & \multicolumn{1}{c}{$E_{B-V}^\dagger$}\\
\multicolumn{1}{c}{$ $} & \multicolumn{1}{c}{($^{\mathrm{h}}\;^{\mathrm{m}}\;^{\mathrm{s}}$)} & \multicolumn{1}{c}{($^\circ\;'\;''$)} & \multicolumn{1}{c}{(deg)} & \multicolumn{1}{c}{(deg)} & \multicolumn{1}{c}{(mag)}\\
\hline
Abell 2744       & $00~14~21.2$ & $-30~23~50.1$ &   $8.89757$ & $-81.23860$ & $0.012$\\
UKIDSS UDS       & $02~17~37.5$ & $-05~12~00.0$ & $158.95220$ & $-51.54158$ & $0.020$\\
Abell 370        & $02~39~52.9$ & $-01~34~36.5$ & $173.00410$ & $-53.57030$ & $0.028$\\
HUDF/GOODS-S     & $03~32~29.5$ & $-27~48~18.3$ & $223.56903$ & $-54.43162$ & $0.007$\\
MACS J0416--2403 & $04~16~08.9$ & $-24~04~28.7$ & $221.08660$ & $-44.05440$ & $0.036$\\
MACS J0717+3745  & $07~17~34.0$ & $+37~44~49.0$ & $180.24429$ & $+21.04515$ & $0.068$\\
COSMOS           & $10~00~27.9$ & $+02~12~03.5$ & $236.82544$ & $+42.11648$ & $0.016$\\
MACS J1149+2223  & $11~49~36.3$ & $+22~23~58.1$ & $228.16350$ & $+75.19849$ & $0.020$\\
HDFN/GOODS-N     & $12~36~54.9$ & $+62~14~18.9$ & $125.86574$ & $+54.80702$ & $0.011$\\
EGS              & $14~19~18.0$ & $+03~55~18.0$ & $103.29148$ & $+54.82531$ & $0.028$\\
Abell S1063      & $22~48~44.4$ & $-44~31~48.5$ & $349.48345$ & $-59.93298$ & $0.010$\\
\hline
\multicolumn{6}{l}{$^\dagger$Taken from the NASA Extragalactic Database based on \citet{sf11}.}
\end{tabular*}
\end{table*}

\section {Modeling {\it Pencil-Beam} Surveys} \label{sec:surveys}

The majority of recent surveys for high-redshift galaxies are based on
deep, narrow-angle imaging observations. The number of stars $N$
within such fields is given by integrating their density distribution 
$p(\bvec{r},M)$ over the sampled volume.  Here we make the usual assumption
that the stellar luminosity function $\phi(M)$ is independent of spatial 
distribution $n(\bvec{r})$ or position in the Galaxy \citep{bahc86}; 
hence
\begin{equation}\label{eqn:dens}
p(\bvec{r},M)\,\dif^3\bvec{r}\,\dif M = n(\bvec{r})\,\phi(M)\,\dif^3\bvec{r}\,\dif{M},
\end{equation}
but we recognize that the cooling of the brown dwarfs may lead to a
vertical dependence on the luminosity function \citep{burg04,ryan11}.
We take $n(\bvec{r})\,\dif^3\bvec{r}$ and
$\phi(M)\,\dif{M}$ to be dimensionless, and adopt the local luminosity
function (as discussed in \S~\ref{sec:lums}).  Therefore we must
renormalize these distributions as:
\begin{equation}
n(\bvec{r_\odot})=\int \Phi(M')\,\dif{M'}\equiv n_{\odot},
\end{equation}
where  $\Phi(M)$  is  the   local  luminosity  function  in  units  of
pc$^{-3}$~mag$^{-1}$.  We renormalize the luminosity function as
$\phi(M)\,\dif{M}\!=\!n_\odot^{-1}\,\Phi(M)\,\dif{M}$ to have units 
of mag$^{-1}$.

\begin{figure}
\epsscale{1.2}
\plotone{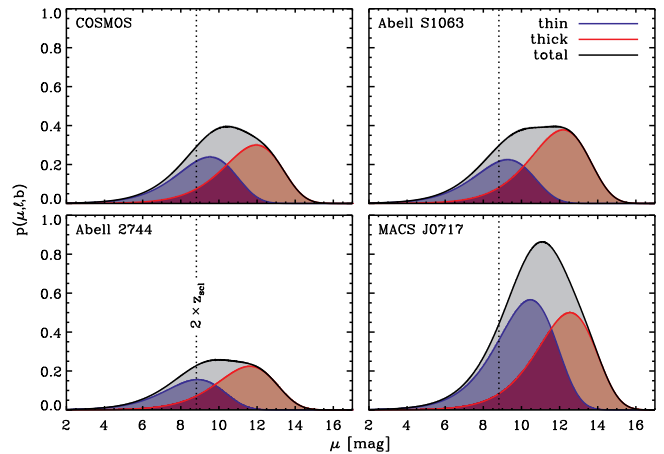}
\caption{Distance distributions for  representative fields.  The blue,
  red, and  gray distributions  show the thin,  thick, and  total disk
  contributions.   For fields  pointed perpendicular  to  the Galactic
  plane,  the distance modulus  distribution from  the thin  disk will
  peak at $2\,z_{\rm scl}$, and the peak is only slightly more distant
  when  including the  thick disk.   It  is important  to note,  these
  distributions  only  depend  on  the  Galactic  model  and  not  the
  luminosity of  the stars in question.  As  such, these distributions
  are generically  true and independent  of bandpass, but  are derived
  under the zero-extinction assumption.  Including Galactic extinction
  is       only      necessary      for       low-latitude      fields
  ($|b|\!\lesssim\!10^\circ$,   though   there   is   a   longitudinal
  dependence  on   this  limit)   and  will  introduce   a  wavelength
  dependence.  \label{fig:pmu}}
\end{figure}

For this  work, we consider  only disk distributions, as  the Galactic
halo is  at least  11~Gyr old \citep{kali}  and any brown  dwarfs here
would  have cooled  below  our spectral-type  range,  but discuss  the
contribution   in   \S~\ref{sec:discuss}.    Assuming   the   standard
double-exponential model, we have
\begin{equation}
\tilde{n}(r,z;r_s,z_s)=n_{\odot}\,\exp\left(\frac{r_\odot-r}{r_s}+\frac{|z_\odot|-|z|}{z_s}\right),
\end{equation}
where $(r_\odot,z_\odot)$ represent the  Solar position in the Galaxy.
So, for a thick and thin disk the total density is given by:
\begin{equation}
n(r,z)=\tilde{n}(r,z;r_s,z_s)+f_t\,\tilde{n}(r,z;R_s,Z_s)
\end{equation}
where $f_t$ is the fraction  of thick disk stars.  The conversion from
these  cylindrical coordinates, assuming  azimuthal symmetry  based on
the Galactic center, to spherical coordinates with to the Sun is given
by \citet{juric08}:
\begin{eqnarray}
r&=&\sqrt{r_0^2+R^2\cos^2 b-2Rr_0\cos b\cos \ell}\\
z&=&R\sin b+z_0,
\end{eqnarray}
\begin{table}
\caption{Galactic Model Parameters}
\label{tab:params}
\begin{tabular*}{0.48\textwidth}
  {@{\extracolsep{\fill}}lccl}
\hline\hline
\multicolumn{1}{c}{name} & \multicolumn{1}{c}{variable} & \multicolumn{1}{c}{value} & \multicolumn{1}{c}{reference}\\
\hline
Solar radius & $r_{\odot}$ & $8000$~pc & \citet{reid93b}\\
Solar height & $z_{\odot}$ &  $25$~pc & \citet{juric08} \\
thin-disk scale length & $r_{scl}$ & $2500$~pc & \citet{juric08} \\
thin-disk scale height & $z_{scl}$ & $290$~pc & \citet{ryan11}\\
thick-disk scale length & $R_{scl}$ & $3600$~pc & \citet{juric08} \\
thick-disk scale height & $Z_{scl}$ & $1000$~pc & \citet{juric08} \\
fraction of thick-disk & $f$ & $0.13$ & \citet{juric08} \\
\hline
\end{tabular*}
\end{table}

where $R$ is the heliocentric distance and $(\ell,b)$ are the Galactic
coordinates of  the field. In  \tab{tab:params} we list  the published
estimates  for the  parameters  of these  distributions  that we  will
adopt.   With a  change  of variables  from  heliocentric distance  to
distance   modulus  $\mu\!=5\log   R  -   5$,  the   distance  modulus
distribution is given by
\begin{equation}\label{eqn:mu}
n(\mu,\ell,b)\,\dif\mu=\Delta\Omega\,n(R,\ell,b)\,R^2\,\dif R
\end{equation}
where $\Delta\Omega$  is the field-of-view.  In  \fig{fig:pmu} we show
the  distance  modulus  distributions  for  the  four  representative,
high-latitude  fields.  The  colors  represent the  thin disk  (blue),
thick disk (red), and total (black).  For fields at very high latitude
(such as  Abell 2744), the typical  distance for a  star is $2\,z_{\rm
  scl}$, which is {\it independent}  of the luminosity of the star and
is generically true  for any population confined to  the Galactic disk
\citep[e.g.][]{green}.  We give the peak and average distance modulus for 
all \nfields~potential \jwst\ deep fields in \tab{tab:pmu}.
\begin{table}
\caption{Distance Modulus Results}
\label{tab:pmu}
\begin{tabular*}{0.48\textwidth}
  {@{\extracolsep{\fill}}lccccc}
\hline\hline
&\multicolumn{2}{c}{thin disk}&&\multicolumn{2}{c}{thick disk}\\
\cline{2-3} \cline{5-6}\\[-12pt]
\multicolumn{1}{c}{Field} & \multicolumn{1}{c}{$\mu_{\rm peak}$} & \multicolumn{1}{c}{$\mu_{\rm ave}$} & & \multicolumn{1}{c}{$\mu_{\rm peak}$} & \multicolumn{1}{c}{$\mu_{\rm ave}$}\\
 & \multicolumn{1}{c}{(mag)} & \multicolumn{1}{c}{(mag)} &  & \multicolumn{1}{c}{(mag)} & \multicolumn{1}{c}{(mag)}\\
\hline
          Abell~2744 &  8.88 &  8.30 & & 11.62 & 11.04 \\
          UKIDSS~UDS &  9.15 &  8.58 & & 11.62 & 11.04 \\
           Abell~370 &  9.13 &  8.53 & & 11.57 & 10.99 \\
        HUDF/GOODS-S &  9.13 &  8.55 & & 11.63 & 11.05 \\
    MACS~J0416--2403 &  9.41 &  8.83 & & 11.82 & 11.24 \\
   MACS~J0717$+$3745 & 10.47 &  9.88 & & 12.55 & 11.96 \\
              COSMOS &  9.52 &  8.94 & & 11.96 & 11.37 \\
   MACS~J1149$+$2223 &  8.86 &  8.26 & & 11.47 & 10.88 \\
        HDFN/GOODS-N &  9.11 &  8.56 & & 11.69 & 11.08 \\
                 EGS &  9.22 &  8.62 & & 11.79 & 11.19 \\
         Abell~S1063 &  9.28 &  8.69 & & 12.19 & 11.60 \\
\hline
\end{tabular*}
\end{table}

\begin{figure}
\epsscale{1.2}
\plotone{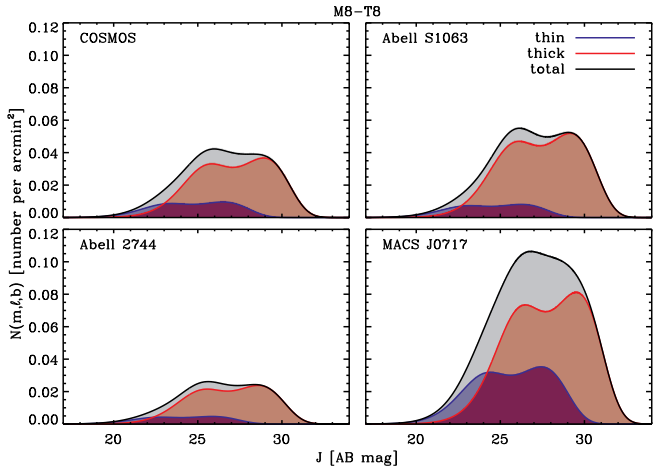}
\caption{Differential  number counts  for  representative \jwst\  deep
  fields.  Here we depict the  contribution from the thin disk (blue),
  thick disk (red), and total disk populations (grey).  The bimodal 
  behavior comes from the two peaks in the extremal ranges of the luminosity 
  function (see \fig{fig:lf}).  In the Appendix, we show similar plots 
  for narrower ranges of spectral type.\label{fig:dif}}
\end{figure}

\begin{figure}
\epsscale{1.2}
\plotone{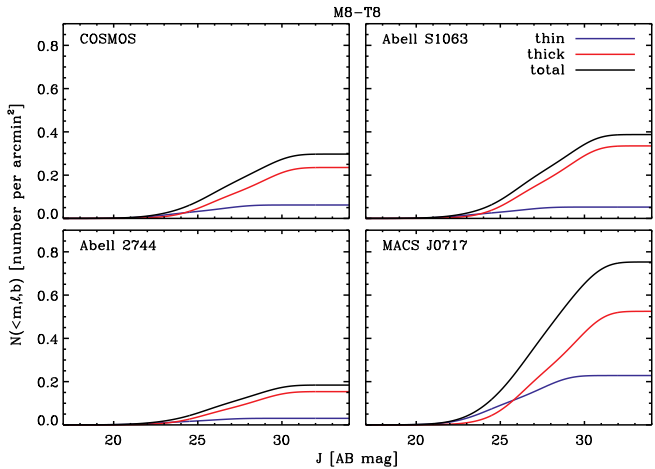}
\caption{Integral number counts for representative \jwst\ deep fields.
  The colors have the same meaning as in \fig{fig:dif}.\label{fig:cum}}
\end{figure}

The   predicted  stellar   number  count   is  given   by  integrating
\eqn{eqn:dens}, which  is the  so-called {\it fundamental  equation of
  stellar statistics} \citep[equation~1 of][]{bahc86}:
\begin{equation}\label{eqn:fess}
N(m,\ell,b)\,\dif m=\Delta\Omega\int\limits_0^\infty n(R,\ell,b)\,\phi(M)\,R^2\,\dif{R}\,\dif{m},
\end{equation}
where  $M$  is  constrained to  satisfy  $M\!=\!m-\mu(R)-A(\ell,b,R)$,
where $A(\cdot)$  is the extinction  along a given line-of-sight  as a
function of  distance. Since the  typical \jwst\ deep field  is likely
situated  far  from the  Galactic  plane,  they  will have  low  color
excesses     \citep[typically     $E_{B-V}\!\lesssim\!0.03$~mag    see
  \tab{tab:fields};][]{sf11}  and   infrared  extinctions  of  $A_{\rm
  IR}\!\lesssim\!0.02$~mag  \citep{sfd}.  Moreover, since  \jwst\ will
be  able to  detect  L0 to  $\sim\!16$~kpc  and T8  to $\sim\!$~3  kpc
\citep[assuming    $J\!\sim\!29$~mag     and    absolute    magnitudes
  of][]{hawl02}, a  full, three-dimensional description  of $A(\cdot)$
is  required  for  reliable number  counts  \citep{sale,ryan11,green}.
Therefore  we adopt  $A(\ell,b,R)\!=\!0$~mag, but  stress  that fields
near the  Galactic plane may  require a more  sophisticated treatment,
which  will  modify  our   predictions.   Using  the  definition  from
\eqn{eqn:mu},  we find  that  \eqn{eqn:fess} is  simply a  convolution
between the  normalized luminosity  function and the  distance modulus
distribution:
\begin{equation}\label{eqn:conv}
N(m,\ell,b)\,\dif m=\Delta\Omega\int\limits_{-\infty}^{+\infty} n(\mu,\ell,b)\,\phi(m-\mu)\,\dif{\mu}\,\dif{m}.
\end{equation}
We show the  differential (\fig{fig:dif}) and integral (\fig{fig:cum})
for  our  representative fields  in  units  of  arcmin$^{-2}$.  As  in
\fig{fig:pmu},  the colors  represent  thin (blue),  thick (red),  and
total (black)  disk components.  In the Appendix,  we tabulate various
statistics of the number counts for the \nfields~potential \jwst\ deep
fields broken down by spectral type (Tables~4--7).  Since stars in the
Milky Way are predominantly confined to an exponential disk, they are
not generally  not found at arbitrarily  faint brightnesses.  Although
the ultracool dwarfs have  very low-luminosities, they are expected to
have       $J\!\sim\!24$~mag       in       high-latitude       fields
($|b|\!\gtrsim\!20^\circ$).

\section{Discussion}\label{sec:discuss}

We  have  demonstrated that  for  an  idealized  survey (ie.~one  with
perfect sample completeness), the  ultracool dwarfs are expected to be
reasonably bright ---  reaching 50\% of their totals  near where their
counts peak of $J\!\sim\!24$~AB  mag.  However a realistic survey will
not   be  so   complete,  and   will  therefore   the   expression  in
\eqn{eqn:fess} should be modified  to include a completeness function.
In most  cases, the  completeness function does  not have  an explicit
dependence  on sky  position,  but  rather is  only  modulated by  the
non-uniformity   of  the   survey   \citep{ryan11,benne}.   For   most
high-redshift  galaxy surveys,  the completeness  is typically  a weak
function    of    brightness    for    ${\rm    AB}\!\lesssim\!25$~mag
\citep[e.g.][]{bou07}, which leaves many  of our results regarding the
surface densities unchanged.

Thus  far we  have explicitly  omitted  a halo  component, since  the
Galactic  halo has an  age of  $\gtrsim\!11$~Gyr \citep{kali}  --- but
here we illustrate  its potential effect on our  results.  To have the
colors  consistent with  a  high-redshift galaxy,  an ultracool  dwarf
would have  a spectral type roughly L5--T5  (see \fig{fig:colors}) and
hence  a temperature of  $2000-1200$~K \citep[Figure  7 of][]{kirk05}.
Based        on        the        expected       cooling        curves
\citep{burrows97,chabrier00,allard01,baraffe03}, a star with an age of
11.4~Gyr  and   temperature  $1200-2000$~K   would  have  a   mass  of
$0.075-0.08~M_{\odot}$ and  likely sustain hydrogen  fusion. Therefore
we include  a Galactic halo following the  parameterization and values
of  \citet{sesar11},  but  calculation  over this  narrow  temperature
range.  The  halo distribution  peaks around $J\!\sim\!30$~AB  mag and
increases  the  total  surface density  of  $\sim\!0.1$~arcmin$^{-2}$.
This   estimate   depends  strongly   on   the   cooling  models   and
star-formation history of  brown dwarfs, and so should  be regarded as
preliminary.   Nevertheless, \jwst\  will have  the ability  to easily
detect early-T dwarfs in the  Galactic halo, indeed the existence of a
sizable population veritably associated with the halo will be a useful
boundary condition on the cooling models.

Throughout this  work we have  tacitly assumed that none  of ultracool
dwarfs  are   in  unresolved  binary  systems.   If   we  assume  that
equal-mass, unresolved binaries  would have same colors, luminosities,
and Galactic  distribution as  single stars, then  the stars  would be
$2.5\log  2\!=\!0.75$ mag  too bright.   If all  stars were  in binary
pairs, then the number counts would shift $\sim\!0.75$~mag to brighter
limits,  but the  overall number  of sources  identified  would remain
constant (since  the binaries  are unresolved).  Certainly  this would
exacerbate  the  issue of  contamination,  however  this {\it  maximum
  binarity} assumption is very  conservative since the observed binary
fractions are more like $\lesssim\!25$\% \citep[e.g.][]{burg07,abe14}.
Moreover, \citet{ryan11}  showed that the  model number counts  do not
significantly change for binary fractions $\leq\!40$\%.

\section{Summary} \label{sec:summary}

Since  brown   dwarfs  have  long  been  recognized   as  a  potential
contaminating     source    in     high-redshift     galaxy    surveys
\citep{yan03,ryan05,cab08,wilkins14},   we   have  provided   tangible
predictions for  the surface density  of these ultracool  dwarfs using
the  best  estimates  for   their  luminosity  function  and  Galactic
distribution.   We find that  the highest  surface densities  are (not
surprisingly)     near     the     Galactic     plane     and     with
$\sim\!1$~arcmin$^{-2}$.    Therefore   in   existing  datasets   with
\hst\  the  numbers are  typically  $\lesssim\!1$  for  a single  WFC3
field-of-view (such as  the HFF or HUDF) or  $\sim\!40$ for wide-field
mosaics (such as CANDELS:~UDS, EGS, or COSMOS). With \jwst\ it will be
possible  to obtain  many  shallow fields  (${\rm AB}\!\sim\!27$)  and
search  for  dwarfs  out  to  $\sim\!6$~kpc,  veritably  sampling  the
Galactic halo.   Although conclusively identifying a given  dwarf as a
member of the  halo will require kinematic data,  which will yield new
constraints  on the cooling  models of  ultracool dwarfs  by providing
robust samples of very old, but massive, objects.

\acknowledgments The  authors would like  to thank Harry  Ferguson for
his  thoughts regarding future  deep fields  with \jwst.   Support for
this work  was provided  by NASA through  grant number 13266  from the
Space Telescope  Science Institute, which  is operated by  AURA, Inc.,
under NASA contract NAS 5-26555. This research has benefited from the
SpeX  Prism  Spectral  Libraries,  maintained  by  Adam  Burgasser  at
http://pono.ucsd.edu/~adam/browndwarfs/spexprism.

{\it Facilities:} \facility{JWST (NIRCam)}

\appendix

\section{Results by Spectral Type}\label{sec:types}

\begin{figure}
\epsscale{1.1}
\plottwo{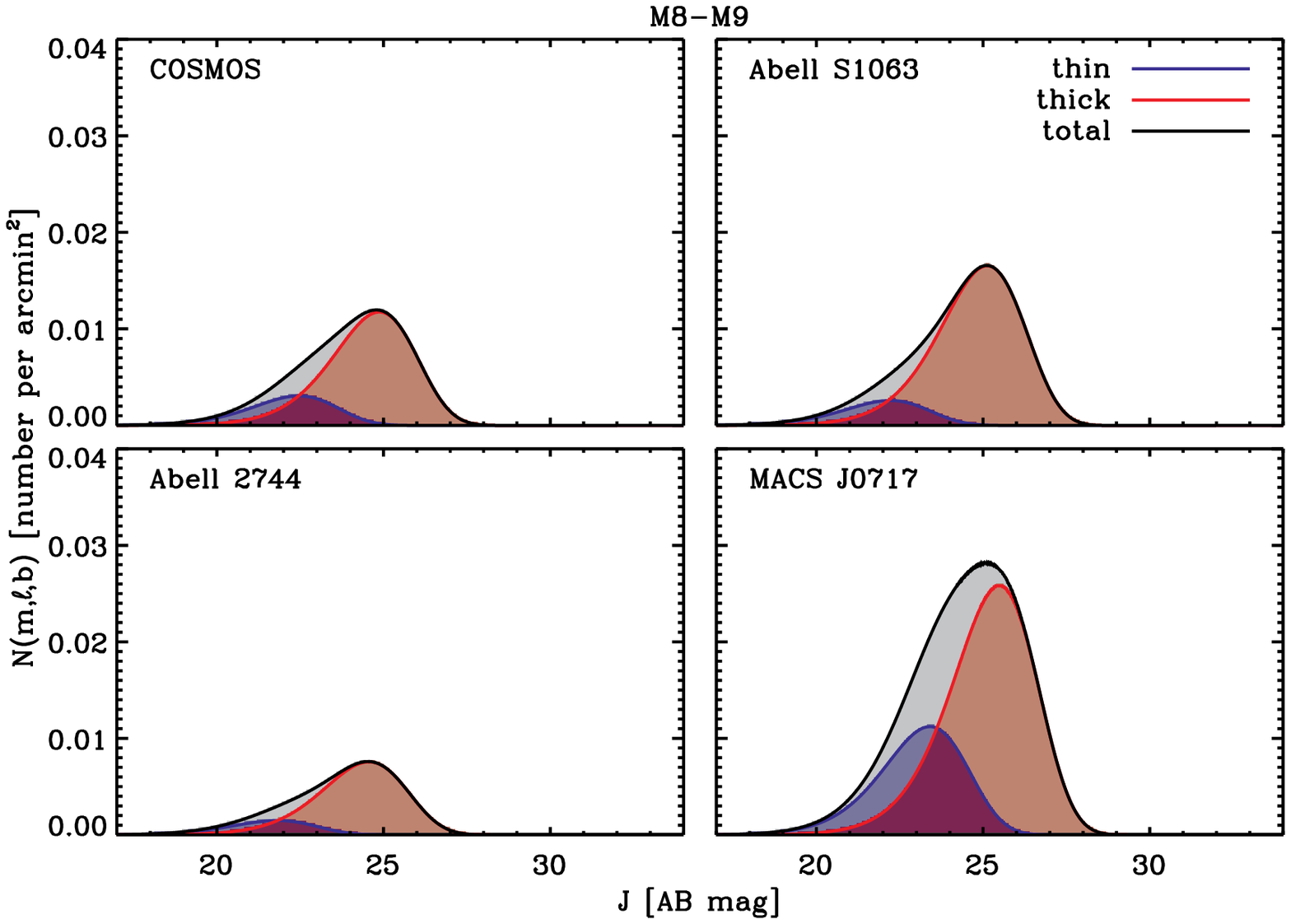}{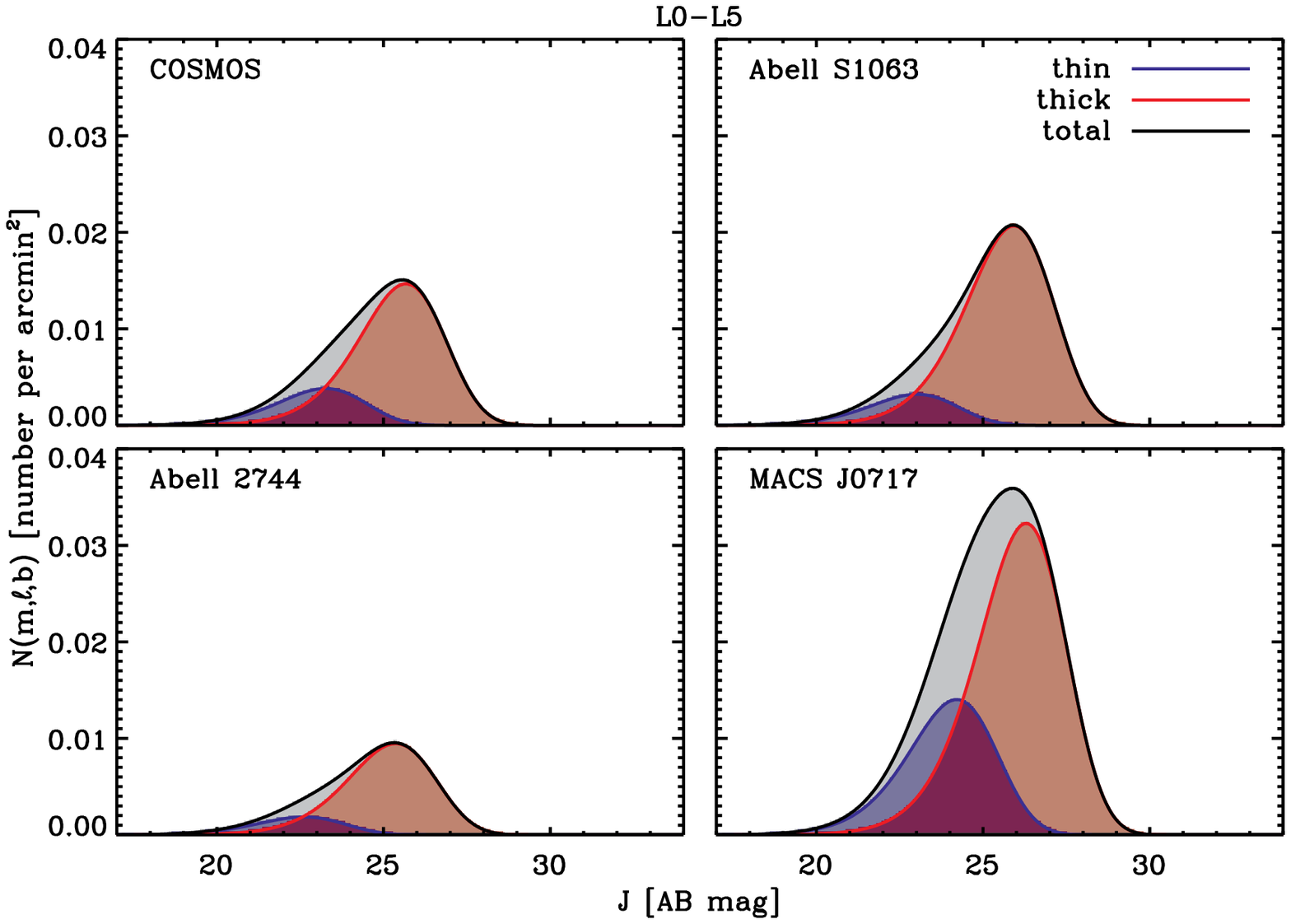}
\plottwo{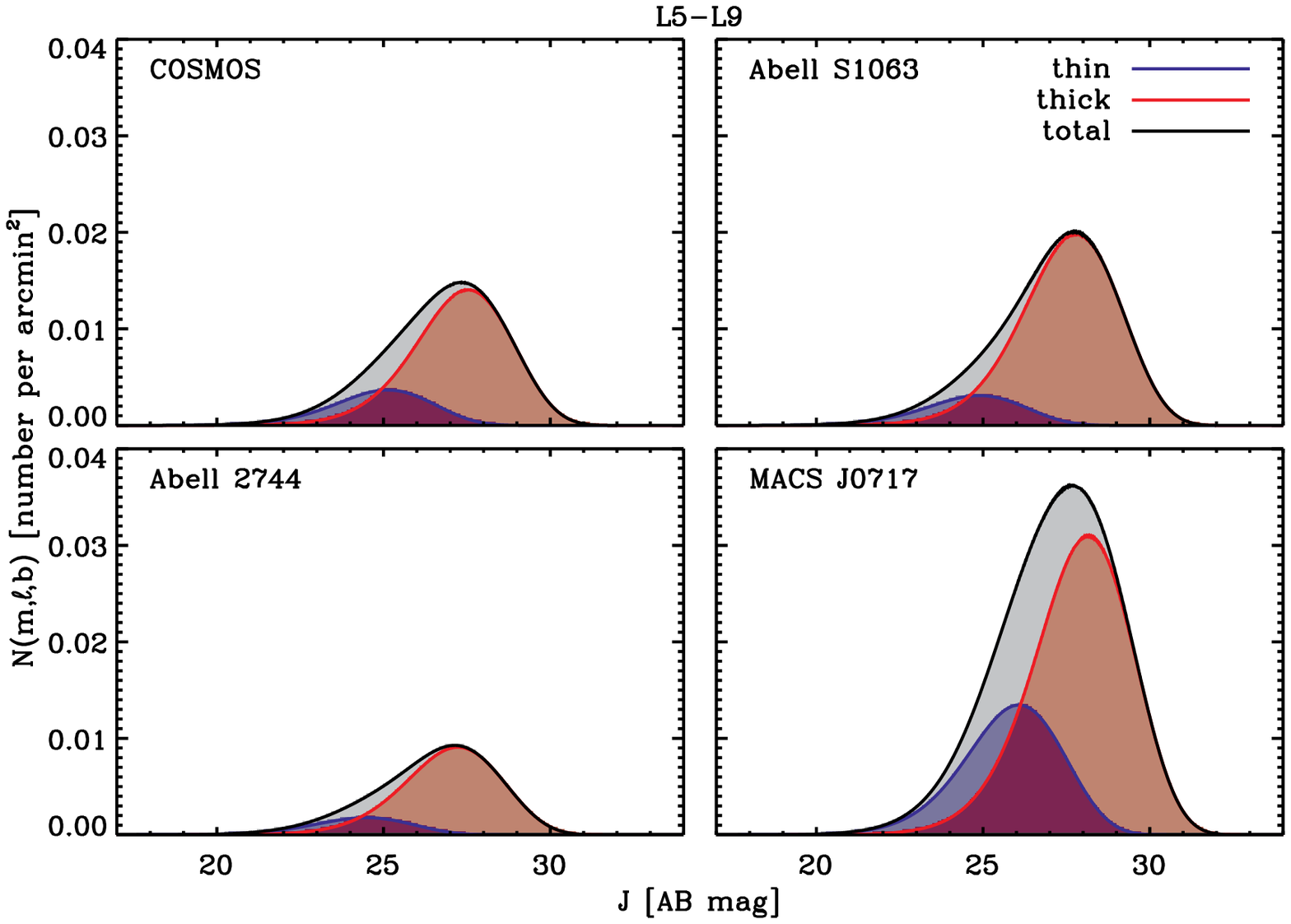}{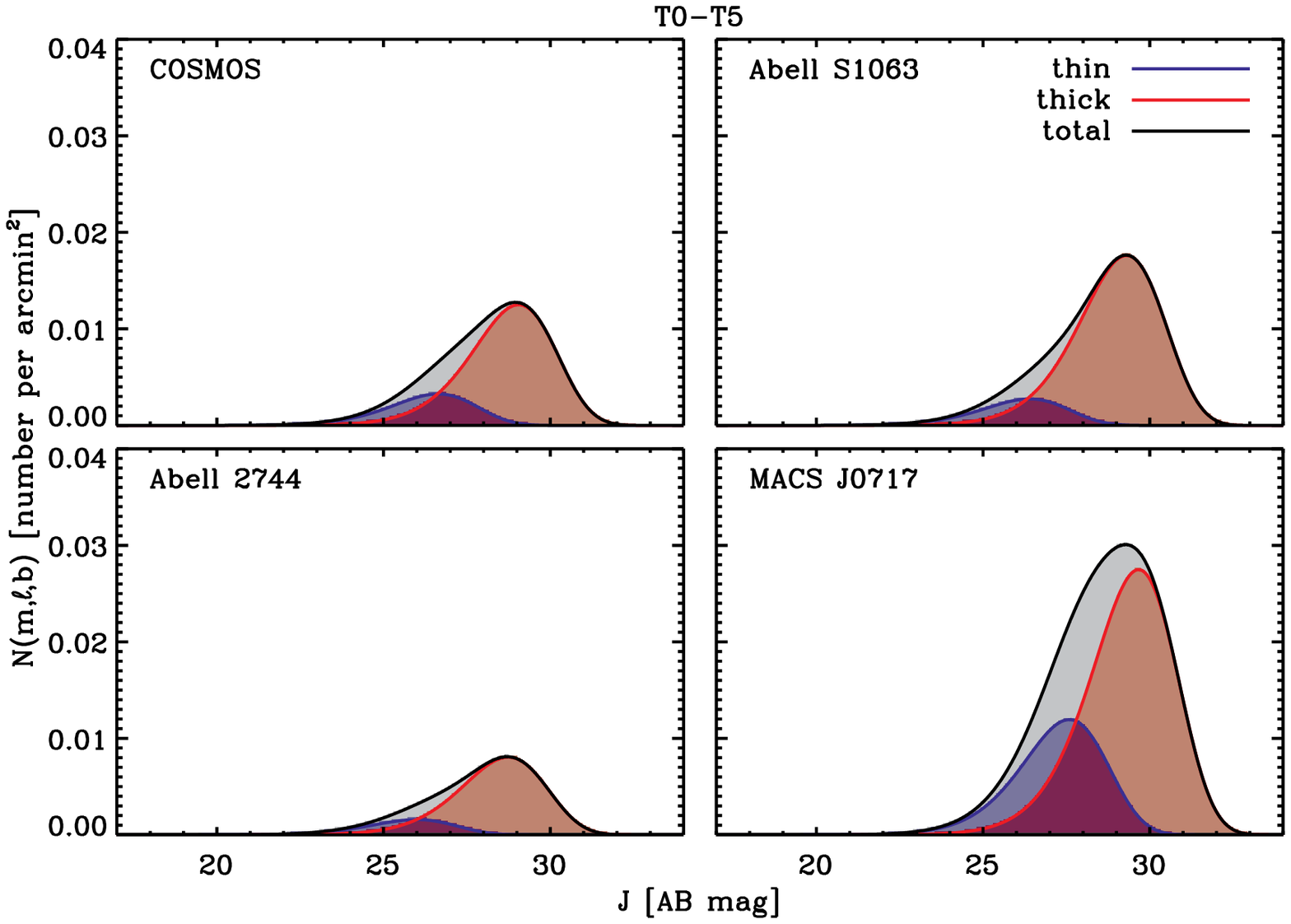}
\caption{Differential  counts for various  spectral type  ranges. Here
  the colors have  the same meaning as \fig{fig:dif},  but we show the
  differential counts  of separate spectral types  M8-M9 (upper left),
  L0-L5  (upper   right),  L5-L9   (lower  left),  and   T0-T5  (lower
  right).\label{fig:dif2}}
\end{figure}

\begin{figure}
\epsscale{1.1}
\plottwo{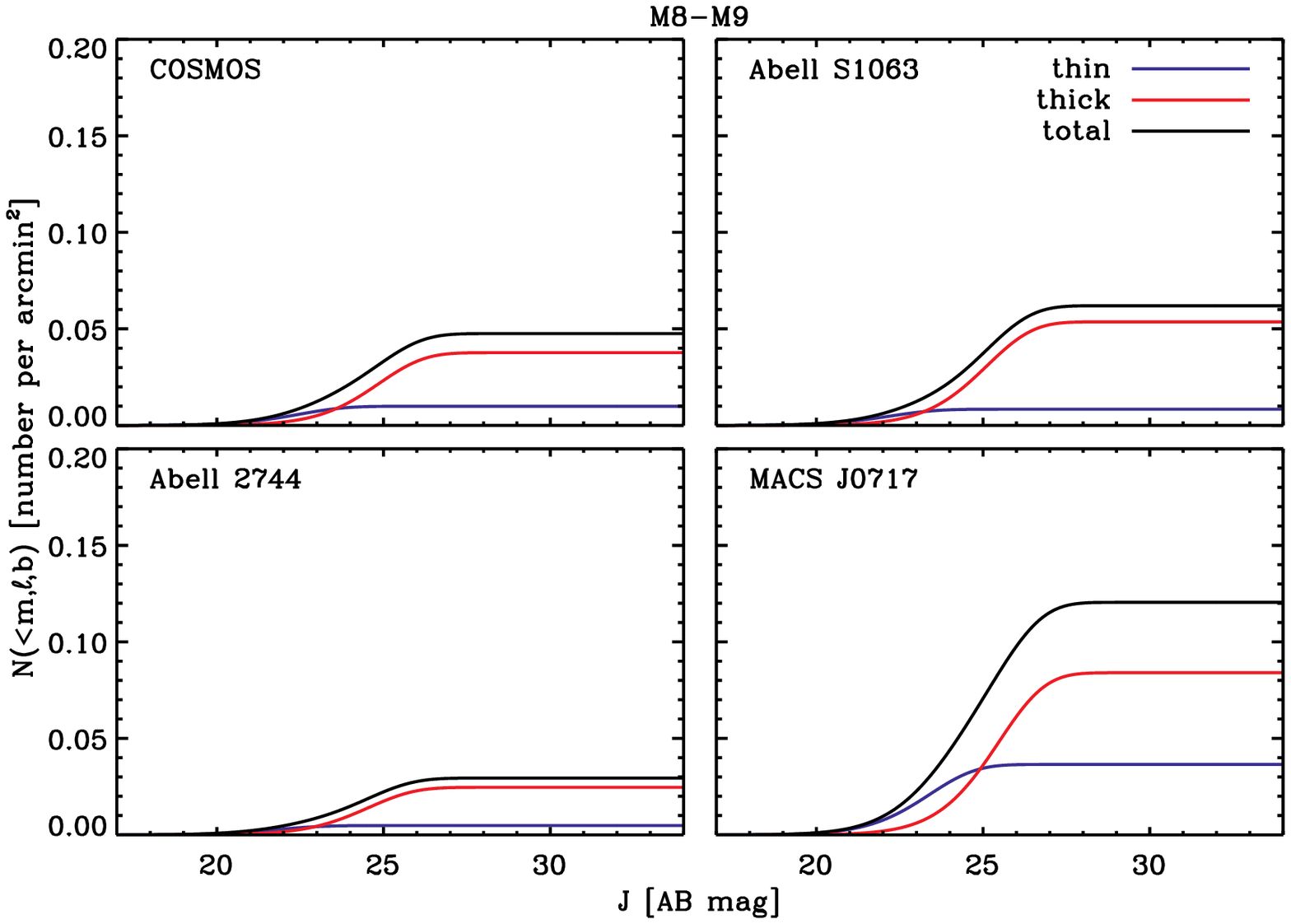}{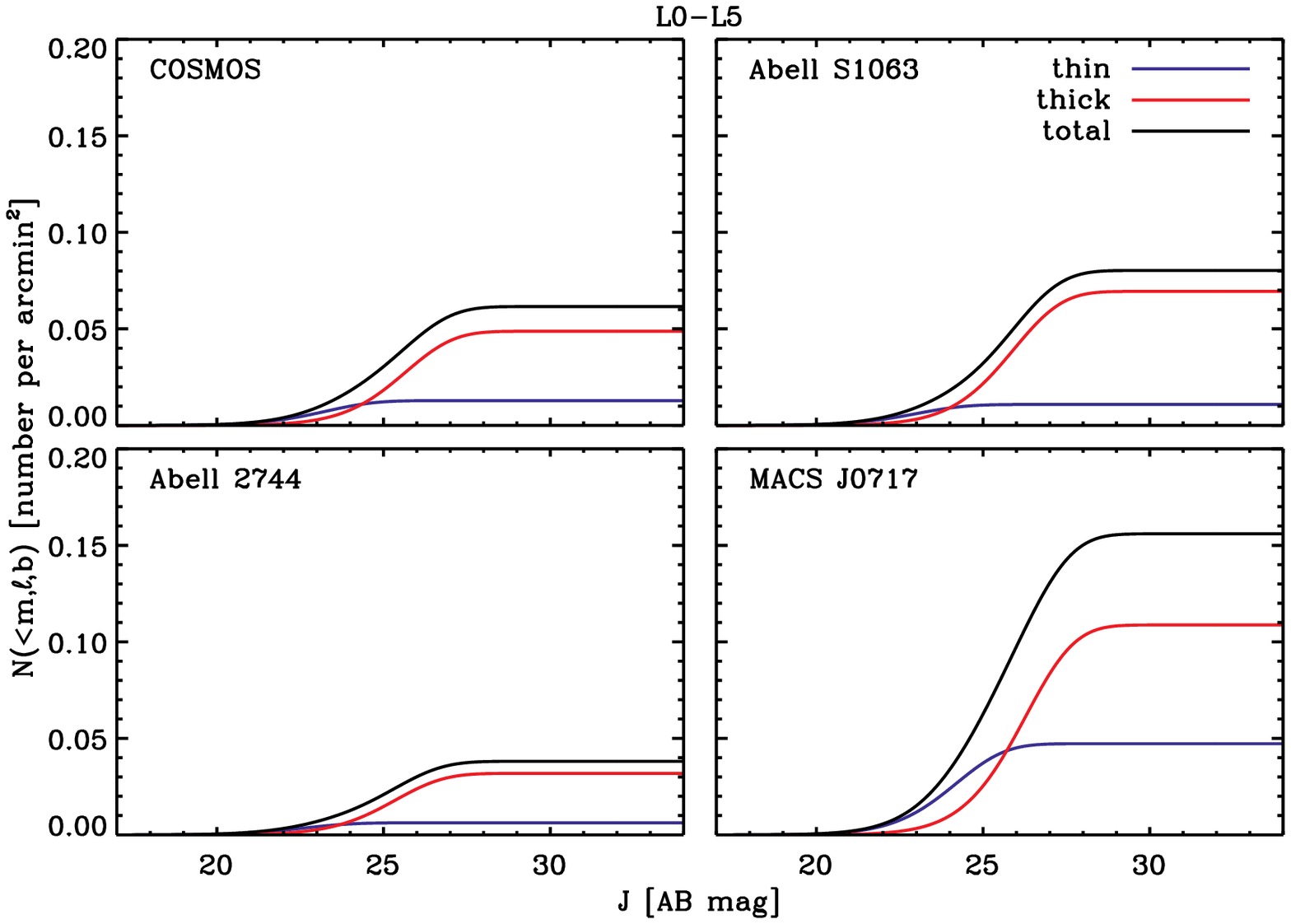}
\plottwo{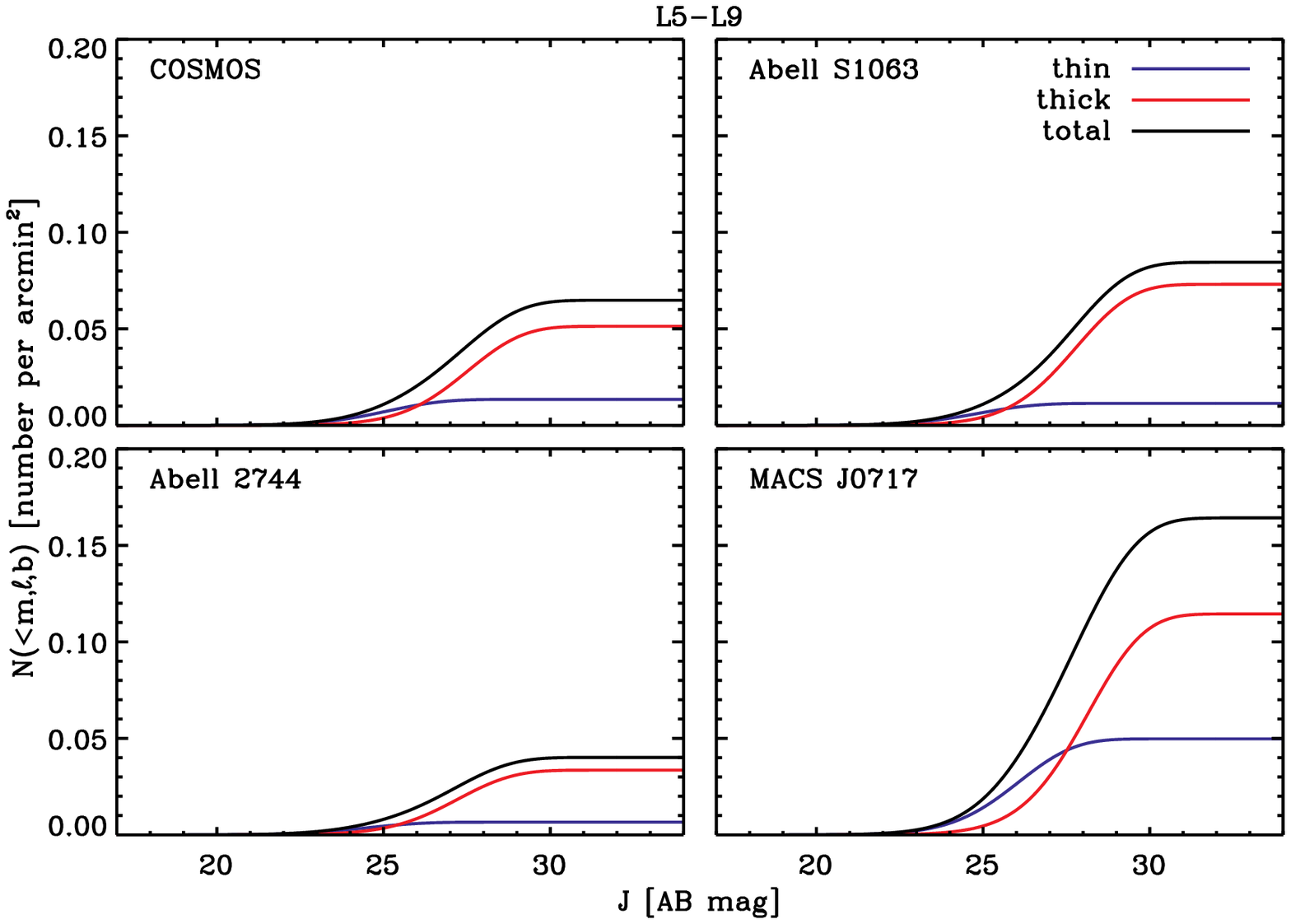}{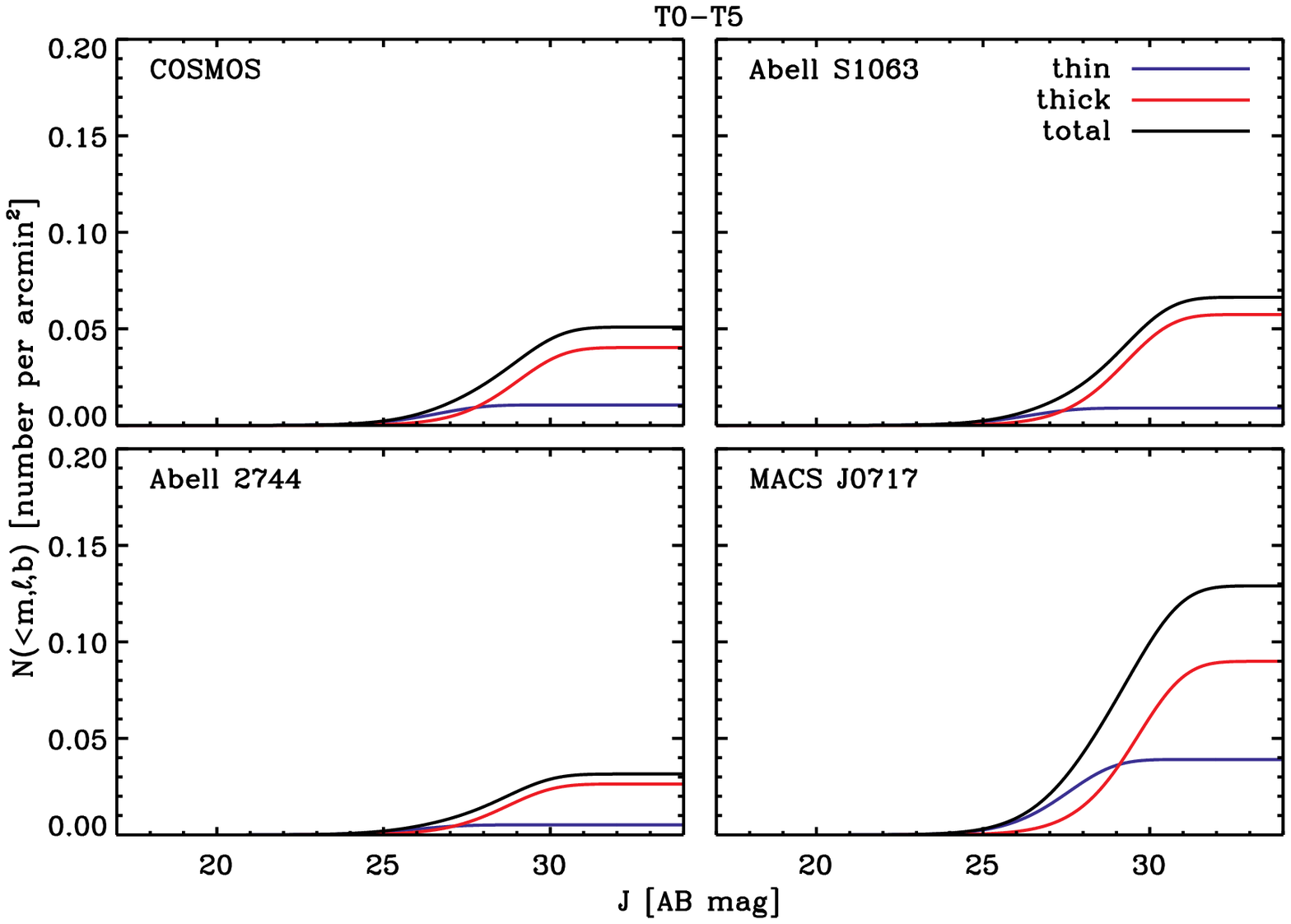}
\caption{Cumulative counts for various spectral type ranges. Here the 
  colors have the same meaning as \fig{fig:cum}, but we show the
  differential counts  of separate spectral types  M8-M9 (upper left),
  L0-L5  (upper   right),  L5-L9   (lower  left),  and   T0-T5  (lower
  right).\label{fig:cum2}}
\end{figure}

In the main text we presented differential and integral counts for all
dwarfs near or  below the hydrogen-burning limit, whose  colors may be
consistent         with        any         high-redshift        galaxy
\citep[ie.~$z\!\gtrsim\!6$][]{yan03,ryan05,cab08,wilkins14}.    However
there  are  numerous   methods  for  selection  high-redshift  samples
(e.g.~photometric redshifts,  color/dropout selection, spectroscopic),
which  will undoubtedly  have  unique selection  biases and  purities.
Therefore  we have  broken  down  the above  results  into four  broad
spectral classes M8-M9 (upper left), L0-L5 (upper right), L5-L9 (lower
left),   and  T0-T5  (lower   right)  in   Figures~\ref{fig:dif2}  and
\ref{fig:cum2}.  As part of the online materials, we freely distribute 
$J$-band number counts for each field and spectral type.

\begin{table*}[h!]
\caption{Number Counts for M8--M9$^\dagger$}
\begin{tabular*}{\textwidth}
  {@{\extracolsep{\fill}}lccccccccccc}
\hline\hline
&\multicolumn{5}{c}{thin disk} & & \multicolumn{5}{c}{thick disk}\\
\cline{2-6} \cline{8-12}\\[-8pt]
\multicolumn{1}{c}{Field} & \multicolumn{1}{c}{$J_{\rm peak}$} & \multicolumn{1}{c}{$J_{1/2}$} & \multicolumn{1}{c}{$\Sigma_{<\!25}$} & \multicolumn{1}{c}{$\Sigma_{<\!27}$} & \multicolumn{1}{c}{$\Sigma_{tot}$} & &\multicolumn{1}{c}{$J_{\rm peak}$} & \multicolumn{1}{c}{$J_{1/2}$} & \multicolumn{1}{c}{$\Sigma_{<\!25}$} & \multicolumn{1}{c}{$\Sigma_{<\!27}$} & \multicolumn{1}{c}{$\Sigma_{tot}$}\\
\hline
          Abell~2744 & 21.82 & 21.57 &  0.48 &  0.48 &  0.48 & &24.56 & 24.31 &  1.74 &  2.45 &  2.46 \\
          UKIDSS~UDS & 22.10 & 21.85 &  0.72 &  0.72 &  0.72 & &24.55 & 24.31 &  1.75 &  2.44 &  2.46 \\
           Abell~370 & 22.04 & 21.80 &  0.66 &  0.66 &  0.66 & &24.51 & 24.26 &  1.66 &  2.29 &  2.30 \\
        HUDF/GOODS-S & 22.07 & 21.82 &  0.69 &  0.69 &  0.69 & &24.56 & 24.31 &  1.78 &  2.50 &  2.52 \\
    MACS~J0416--2403 & 22.35 & 22.09 &  1.00 &  1.01 &  1.01 & &24.75 & 24.50 &  2.15 &  3.26 &  3.30 \\
   MACS~J0717$+$3745 & 23.40 & 23.15 &  3.45 &  3.65 &  3.65 & &25.49 & 25.23 &  3.62 &  7.87 &  8.40 \\
              COSMOS & 22.45 & 22.20 &  0.99 &  0.99 &  0.99 & &24.87 & 24.62 &  2.33 &  3.72 &  3.77 \\
   MACS~J1149$+$2223 & 21.78 & 21.53 &  0.39 &  0.39 &  0.39 & &24.40 & 24.15 &  1.42 &  1.88 &  1.89 \\
        HDFN/GOODS-N & 22.08 & 21.83 &  0.59 &  0.59 &  0.59 & &24.59 & 24.34 &  1.76 &  2.50 &  2.52 \\
                 EGS & 22.14 & 21.89 &  0.64 &  0.64 &  0.64 & &24.70 & 24.45 &  1.98 &  2.93 &  2.95 \\
         Abell~S1063 & 22.22 & 21.97 &  0.84 &  0.84 &  0.84 & &25.12 & 24.87 &  2.89 &  5.22 &  5.36 \\
\hline
\multicolumn{12}{l}{$^\dagger$Columns 2--3 and 7--8 are in AB magnitudes while columns 4--6 and 9--11 are in $10^{-2}$~arcmin$^{-2}$}
\end{tabular*}
\end{table*}

\begin{table*}[h!]
\caption{Number Counts for L0--L5$^\dagger$}
\begin{tabular*}{\textwidth}
  {@{\extracolsep{\fill}}lccccccccccc}
\hline\hline
&\multicolumn{5}{c}{thin disk} & & \multicolumn{5}{c}{thick disk}\\
\cline{2-6} \cline{8-12}\\[-8pt]
\multicolumn{1}{c}{Field} & \multicolumn{1}{c}{$J_{\rm peak}$} & \multicolumn{1}{c}{$J_{1/2}$} & \multicolumn{1}{c}{$\Sigma_{<\!25}$} & \multicolumn{1}{c}{$\Sigma_{<\!27}$} & \multicolumn{1}{c}{$\Sigma_{tot}$} & &\multicolumn{1}{c}{$J_{\rm peak}$} & \multicolumn{1}{c}{$J_{1/2}$} & \multicolumn{1}{c}{$\Sigma_{<\!25}$} & \multicolumn{1}{c}{$\Sigma_{<\!27}$} & \multicolumn{1}{c}{$\Sigma_{tot}$}\\
\hline
          Abell~2744 & 22.62 & 22.39 &  0.62 &  0.63 &  0.63 & &25.36 & 25.13 &  1.47 &  2.99 &  3.19 \\
          UKIDSS~UDS & 22.91 & 22.68 &  0.91 &  0.93 &  0.93 & &25.36 & 25.13 &  1.47 &  2.99 &  3.18 \\
           Abell~370 & 22.85 & 22.62 &  0.84 &  0.86 &  0.86 & &25.31 & 25.08 &  1.41 &  2.81 &  2.97 \\
        HUDF/GOODS-S & 22.87 & 22.64 &  0.87 &  0.89 &  0.89 & &25.37 & 25.14 &  1.50 &  3.06 &  3.26 \\
    MACS~J0416--2403 & 23.15 & 22.92 &  1.25 &  1.30 &  1.30 & &25.56 & 25.33 &  1.73 &  3.91 &  4.27 \\
   MACS~J0717$+$3745 & 24.21 & 23.98 &  3.72 &  4.71 &  4.72 & &26.29 & 26.06 &  2.54 &  8.35 & 10.87 \\
              COSMOS & 23.26 & 23.03 &  1.22 &  1.28 &  1.28 & &25.67 & 25.44 &  1.83 &  4.39 &  4.88 \\
   MACS~J1149$+$2223 & 22.58 & 22.36 &  0.50 &  0.50 &  0.50 & &25.20 & 24.98 &  1.24 &  2.34 &  2.45 \\
        HDFN/GOODS-N & 22.88 & 22.66 &  0.75 &  0.76 &  0.76 & &25.40 & 25.17 &  1.47 &  3.05 &  3.26 \\
                 EGS & 22.94 & 22.71 &  0.80 &  0.83 &  0.83 & &25.50 & 25.27 &  1.61 &  3.54 &  3.82 \\
         Abell~S1063 & 23.02 & 22.79 &  1.05 &  1.08 &  1.08 & &25.92 & 25.69 &  2.17 &  5.92 &  6.94 \\
\hline
\multicolumn{12}{l}{$^\dagger$Columns 2--3 and 7--8 are in AB magnitudes while columns 4--6 and 9--11 are in $10^{-2}$~arcmin$^{-2}$}
\end{tabular*}
\end{table*}

\begin{table*}[h!]
\caption{Number Counts for L5--L9$^\dagger$}
\begin{tabular*}{\textwidth}
  {@{\extracolsep{\fill}}lccccccccccc}
\hline\hline
&\multicolumn{5}{c}{thin disk} & & \multicolumn{5}{c}{thick disk}\\
\cline{2-6} \cline{8-12}\\[-8pt]
\multicolumn{1}{c}{Field} & \multicolumn{1}{c}{$J_{\rm peak}$} & \multicolumn{1}{c}{$J_{1/2}$} & \multicolumn{1}{c}{$\Sigma_{<\!25}$} & \multicolumn{1}{c}{$\Sigma_{<\!27}$} & \multicolumn{1}{c}{$\Sigma_{tot}$} & &\multicolumn{1}{c}{$J_{\rm peak}$} & \multicolumn{1}{c}{$J_{1/2}$} & \multicolumn{1}{c}{$\Sigma_{<\!25}$} & \multicolumn{1}{c}{$\Sigma_{<\!27}$} & \multicolumn{1}{c}{$\Sigma_{tot}$}\\
\hline
          Abell~2744 & 24.49 & 24.30 &  0.45 &  0.65 &  0.66 & &27.23 & 27.04 &  0.35 &  1.64 &  3.35 \\
          UKIDSS~UDS & 24.77 & 24.58 &  0.60 &  0.95 &  0.98 & &27.22 & 27.03 &  0.35 &  1.64 &  3.35 \\
           Abell~370 & 24.72 & 24.53 &  0.57 &  0.88 &  0.91 & &27.18 & 26.99 &  0.34 &  1.58 &  3.13 \\
        HUDF/GOODS-S & 24.74 & 24.55 &  0.58 &  0.91 &  0.94 & &27.23 & 27.04 &  0.36 &  1.68 &  3.43 \\
    MACS~J0416--2403 & 25.01 & 24.82 &  0.75 &  1.30 &  1.37 & &27.42 & 27.23 &  0.39 &  1.97 &  4.49 \\
   MACS~J0717$+$3745 & 26.07 & 25.88 &  1.41 &  3.91 &  4.97 & &28.15 & 27.96 &  0.46 &  3.06 & 11.45 \\
              COSMOS & 25.12 & 24.93 &  0.70 &  1.27 &  1.35 & &27.54 & 27.35 &  0.39 &  2.10 &  5.13 \\
   MACS~J1149$+$2223 & 24.45 & 24.26 &  0.37 &  0.52 &  0.53 & &27.07 & 26.88 &  0.31 &  1.37 &  2.58 \\
        HDFN/GOODS-N & 24.75 & 24.56 &  0.50 &  0.78 &  0.80 & &27.26 & 27.07 &  0.35 &  1.65 &  3.43 \\
                 EGS & 24.81 & 24.62 &  0.53 &  0.84 &  0.87 & &27.36 & 27.17 &  0.36 &  1.83 &  4.03 \\
         Abell~S1063 & 24.88 & 24.69 &  0.67 &  1.10 &  1.14 & &27.79 & 27.60 &  0.43 &  2.54 &  7.31 \\
\hline
\multicolumn{12}{l}{$^\dagger$Columns 2--3 and 7--8 are in AB magnitudes while columns 4--6 and 9--11 are in $10^{-2}$~arcmin$^{-2}$}
\end{tabular*}
\end{table*}

\begin{table*}[h!]
\caption{Number Counts for T0--T5$^\dagger$}
\begin{tabular*}{\textwidth}
  {@{\extracolsep{\fill}}lccccccccccc}
\hline\hline
&\multicolumn{5}{c}{thin disk} & & \multicolumn{5}{c}{thick disk}\\
\cline{2-6} \cline{8-12}\\[-8pt]
\multicolumn{1}{c}{Field} & \multicolumn{1}{c}{$J_{\rm peak}$} & \multicolumn{1}{c}{$J_{1/2}$} & \multicolumn{1}{c}{$\Sigma_{<\!25}$} & \multicolumn{1}{c}{$\Sigma_{<\!27}$} & \multicolumn{1}{c}{$\Sigma_{tot}$} & &\multicolumn{1}{c}{$J_{\rm peak}$} & \multicolumn{1}{c}{$J_{1/2}$} & \multicolumn{1}{c}{$\Sigma_{<\!25}$} & \multicolumn{1}{c}{$\Sigma_{<\!27}$} & \multicolumn{1}{c}{$\Sigma_{tot}$}\\
\hline
          Abell~2744 & 25.99 & 25.75 &  0.15 &  0.44 &  0.52 & &28.74 & 28.49 &  0.05 &  0.41 &  2.63 \\
          UKIDSS~UDS & 26.28 & 26.03 &  0.18 &  0.60 &  0.77 & &28.73 & 28.49 &  0.05 &  0.41 &  2.63 \\
           Abell~370 & 26.22 & 25.98 &  0.17 &  0.57 &  0.71 & &28.69 & 28.44 &  0.05 &  0.40 &  2.46 \\
        HUDF/GOODS-S & 26.25 & 26.00 &  0.18 &  0.58 &  0.74 & &28.74 & 28.49 &  0.05 &  0.41 &  2.69 \\
    MACS~J0416--2403 & 26.52 & 26.27 &  0.20 &  0.77 &  1.08 & &28.93 & 28.68 &  0.05 &  0.45 &  3.53 \\
   MACS~J0717$+$3745 & 27.58 & 27.34 &  0.26 &  1.57 &  3.91 & &29.66 & 29.42 &  0.05 &  0.54 &  8.99 \\
              COSMOS & 26.63 & 26.38 &  0.18 &  0.73 &  1.06 & &29.05 & 28.80 &  0.05 &  0.46 &  4.03 \\
   MACS~J1149$+$2223 & 25.96 & 25.71 &  0.13 &  0.36 &  0.41 & &28.58 & 28.33 &  0.04 &  0.36 &  2.02 \\
        HDFN/GOODS-N & 26.26 & 26.01 &  0.15 &  0.50 &  0.63 & &28.77 & 28.52 &  0.04 &  0.40 &  2.69 \\
                 EGS & 26.32 & 26.07 &  0.16 &  0.53 &  0.68 & &28.88 & 28.63 &  0.05 &  0.42 &  3.16 \\
         Abell~S1063 & 26.39 & 26.15 &  0.19 &  0.68 &  0.90 & &29.30 & 29.05 &  0.05 &  0.51 &  5.74 \\
\hline
\multicolumn{12}{l}{$^\dagger$Columns 2--3 and 7--8 are in AB magnitudes while columns 4--6 and 9--11 are in $10^{-2}$~arcmin$^{-2}$}
\end{tabular*}
\end{table*}

\end{document}